\documentclass[12pt,english, preprint]{elsarticle}

\newcommand{\Journal}[4]{{#1} #4 {\bf #2}, #3 }
\usepackage{graphicx}
\usepackage[latin1]{inputenc}
\graphicspath{{.//}{figures//}}
\usepackage{lineno}
\usepackage{chngpage}
\usepackage{float}
\usepackage{floatflt}
\usepackage{amssymb}
\usepackage{amsmath}
\usepackage{color}
\usepackage{fancybox}
\usepackage{moreverb}
\usepackage{wrapfig}
\usepackage[german]{varioref}
\usepackage{picins}
\usepackage{listings}
\usepackage{scrpage2}
\usepackage{geometry}
\usepackage{multirow}

	\restylefloat{figure,table,listing}

\newcommand{\NIMA}{{\em Nucl. Instrum. Methods} A}

\newcommand{\NPA}{{\em Nucl. Phys.} A}
\newcommand{\PLB}{{\em Phys. Lett.}  {\bf B}}
\newcommand{\PRL}{\em Phys. Rev. Lett.}
\newcommand{\PRD}{{\em Phys. Rev.} {\bf D}}
\newcommand{\PRC}{{\em Phys. Rev.} C}

\newcommand{\ZPA}{{\em Z. Phys.} A}

\newcommand{\PAN}{\em Phys. Atom. Nucl.}

\newcommand{\JPCS}{\em J. Phys. Conf. Ser.}
\newcommand{\JINST}{\em JINST}
\newcommand{\RoSI}{\em Rev. of Scientific Instruments}
\newcommand{\APPH}{\em Astroparticle Physics}
\newcommand{\EPL}{\em Europhys. Lett.}
\newcommand{\EPJA}{\em Euro. Phys. J. A}

\newcommand{\CPC}{\em Chin. Phys.  C}

\newcommand{\JPG}{\em Journal of Physics G}
\newcommand{\NAT}{\em Nature}

\newcommand{\bpbp}{\mbox{$\beta^+\beta^+$} }
\newcommand{\ecec}{\mbox{$ECEC$} }
\newcommand{\bec}{\mbox{$\beta^+EC$} }
\newcommand{\bnel}{\mbox{$\bar{\nu}_e$} }

\newcommand{\obb}{0\mbox{$\nu\beta\beta$ - decay} } 
\newcommand{\zbb}{2\mbox{$\nu\beta\beta$ - decay} }

\newcommand{\nel}{\mbox{$\nu_e$}}

\newcommand{\gess}{\mbox{$^{76}$Ge }}

\newcommand{\teha}{\mbox{$^{128}$Te }}

\newcommand{\tehd}{\mbox{$^{130}$Te }}

\newcommand{\xehs}{\mbox{$^{136}$Xe }} 
\newcommand{\xehv}{\mbox{$^{134}$Xe }} 
\newcommand{\xehf}{\mbox{$^{124}$Xe }} 
\newcommand{\xehz}{\mbox{$^{126}$Xe }} 
\newcommand{\kraf}{\mbox{$^{85}$Kr }}

\newcommand{\tzn}{\mbox{T$_{1/2}^{2\nu}$} }

\newcommand{\be}{\begin{equation}}
\newcommand{\ee}{\end{equation}}
\def\bea{\begin{eqnarray}} 
\def\eea{\end{eqnarray}} 
\newcommand{\ra}{\rightarrow }

\begin{document}
\begin{frontmatter}
\title{Double beta decay searches of \xehv , \xehz and \xehf with large scale Xe detectors}

\author[uop]{N. Barros}
\author[iktp]{J. Thurn}
\author[iktp]{K Zuber}

\address[iktp]{Institut f\"ur Kern- und Teilchenphysik, Technische Universit\"at Dresden, 01069 Dresden, Germany}
\address[uop]{Dept. of Physics and Astronomy, University of Pennsylvania, Philadelphia, PA 19104-6396, USA}
\begin{abstract}
The sensitivity for double beta decay studies of \xehv and \xehf is investigated assuming a potential large scale Xe experiment developed for dark matter searches
depleted in \xehs. The opportunity for an observation of the \zbb of \xehv is explored for various scenarios. A positive observation should be possible for all calculated nuclear matrix elements. The detection of 2$\nu$ ECEC of \xehf can be probed in all scenarios covering the theoretical predicted half-life uncertainties and a potential search for \xehz is discussed. The sensitivity to \bec decay of \xehf is discussed and a positive observation might be possible, while \bpbp decay still remains unobservable. The performed studies take into account solar pp-neutrino interactions, $^{85}$Kr beta decay and remaining \xehs double beta decay as background
components in the depleted detector.
\end{abstract}

\end{frontmatter}

\section{Introduction}
The search for physics beyond the standard model is a wide spread activity in accelerator and non-accelerator physics. Among all the searches for new physics beyond the Standard Model total lepton number violation plays an important role. The golden channel to search for total lepton number violation is neutrinoless double beta decay 
\be
(Z,A) \ra (Z+2,A) + 2 e^-  \quad (\obb).
\ee   
Any Beyond Standard Model (BSM) physics allowing $\Delta L =2$ processes can contribute to the decay rate. However,
it has been shown that its observation would imply that 
neutrinos are their own antiparticles (Majorana neutrinos) which is an
essential ingredient for leptogenesis, explaining the baryon asymmetry in the universe with the help of Majorana neutrinos (see for example \cite{fon12}). However, unknown is how much individual BSM processes contribute to the double beta decay rate. \\
To observe this process, single beta decay has to be forbidden by energy conservation or at least strongly suppressed. For this reason only 35 potential double beta emitters exist. As the phase space for these decays scales strongly with the Q-value, searches are using only those nuclides with a Q-value above 2 MeV, reducing the list of candidates to 11. Lower limits on half-lives beyond 10$^{25}$ years of the neutrino less mode have been measured for the isotopes \gess and \xehs  \cite{gan13,ago13,alb14}. In addition, the allowed  process of neutrino accompanied double beta decay 
\be
(Z,A) \ra (Z+2,A) + 2 e^-  + 2 \bnel \quad (\zbb)
\ee      
will occur. It is the rarest decay measured in nature and has been observed in more than ten isotopes. However, in both cases the measured quantity is a half-life, which is linked to phase space $G$ and nuclear transition matrix elements $M$. In case of \zbb the matrix element is purely Gamow-Teller (GT) and the relation is 
\be
\left(\tzn\right)^{-1} = G \times \mid M_{GT} \mid^2\,,
\ee
which does not contain any unknowns from the particle physics point of view, as opposed to \obb. Thus, half-life measurements for \zbb will provide valuable information on the nuclear matrix elements which can be directly compared with theory. From the nuclear structure point of view the measurement
is important, as \zbb half-lives for the most studied nuclei have been measured. Furthermore, by comparing double beta emitters from the same element (separated by a difference of two neutrons in the same shell) the matrix element calculations can be well probed which has been done for the Te systems \cite{bes10}.
Nature offers a few elements  with 2(3) emitters:  $^{46,48}$Ca, $^{80,82}$Se, $^{94,96}$Zr, $^{98,100}$Mo, $^{114,116}$Cd, $^{122,124}$Sn, $^{134,136}$Xe and $^{146,148,150}$Nd. However due to the Q$^{11}$-dependence of the phase space for \zbb the one with the highest Q-value makes the remaining ones for the same element more or less unobservable. \\
So far only examples passing this rule are geochemical experiments using ancient Te-ores. Here both \teha and \tehd half-lives could be measured. In this way sensitive matrix element ratios can be computed and compared with the experimental values. As it is impractical to enrich an element in the minor emitter (the one with the lower Q-value) just for this purpose, a potential measurement must be done as a byproduct of other ongoing experiments. In this case we propose that \xehv very likely is the first isotope which will be studied in a laboratory experiment, as this can be done in future large-scale liquid xenon (LXe) detectors used for dark matter searches. \\
Equivalent processes to double beta decay with the emission of two electrons could occur on the right side of the mass parabola of even-even isobars. In this case three different decay modes are possible involving positrons and electron capture (EC)
\begin{eqnarray}
\label{eq:ecec}
(Z,A) &\ra  (Z-2,A) + 2 e^+ \; (+ 2 \nel) \quad & \mbox{(\bpbp)} \label{eqn:b+b+}\\
e^- + (Z,A) &\ra (Z-2,A) + e^+ \; (+ 2 \nel) \quad & \mbox{(\bec)} \label{eqn:ecec}\\
2 e^- + (Z,A) &\ra (Z-2,A) \; (+ 2 \nel) \quad & \mbox{(\ecec)} \label{eqn:ecb+}
\end{eqnarray}
Again 35 potential emitters exist for \ecec decay but only six nuclides have a high enough Q-value beyond 4 m$_ec^2$ to allow for double positron decay, one of them being \xehf. No neutrino accompanied decay of any of those has been observed. 
Furthermore, it has been shown that  the neutrinoless \bec mode has an enhanced sensitivity to potential right-handed weak currents \cite{hir94} and thus 
can help to disentangle the contributions to \obb if ever observed.\\
Nowadays detectors using Xe in different forms are common. EXO-200 and KamLAND-Zen are using enriched \xehs for double beta decay searches. Thus, they are not suited for the investigations described in this paper. However, dark matter detectors, especially those with Xe depleted in \xehs, can do a sensitive search. Currently Xenon-100 \cite{apr12}, LUX \cite{ake14} and XMASS \cite{abe13} are running, with Xenon-1T in the building up phase and LZ in preparation. These will already be tonne scale LXe detectors. Furthermore LZ at its final stage \cite{mal11} considers 20 tonnes of LXe which is also one of the considered options for the DARWIN project \cite{bau11}. Considering these dimensions it is interesting to explore their potential for double beta decay searches. However, issues related to fiducial volume cuts and
the dynamic range of the data acquisition system are ignored in this paper.
\section{Generic study of \xehv}

\subsection{Estimates of nuclides and expected half-lives}
\label{sec:estimates}
The  double beta decay of \xehv into $^{134}$Ba has a Q-value of 825.8 $\pm 0.9$ keV \cite{ame12} and its natural abundance is 10.4 \%. The current limit on the \zbb half-life is $1.1 \times 10^{16}$ yrs \cite{bar89}. In natural Xe-detectors it is very unlikely to observe this process due to the presence of \xehs. Its much higher Q-value of  2457.83 $\pm$ 0.37 keV \cite{red07} will completely cover a potential \xehv signal. However, next generation dark matter experiments like Xenon1T will use LXe depleted in \xehs and thus enhance the probability of an observation. \\
 As the maximum in the sum energy spectrum of the two electrons in \zbb is around 30\% of the Q-value, the region of interest for a search will be around 100-500 keV. 
From the experimental side, the abundances of \xehv, \xehs and \kraf, as well as the energy resolution in this range are the major unknown parameters. The latter is due to the fact that these experiments are normally tuned to optimize dark matter searches in the range of a few keV. \kraf is a common background in low energy experiments which decays by beta emission with a Q value of 687.4 keV.\\ 
 To estimate the amount of \xehv in the depleted Xe, data from enriched Xe experiments were used. EXO-200 quotes abundances of these isotopes in their Xe of  80.6\% \xehs and 19.4 \% \xehv \cite{aug12} while KamLAND-Zen is using 90.93 \% enriched \xehs and 8.89 \% \xehv \cite{gan12} . Taking the KamLAND-Zen enrichment values, three different scenarios are defined in this work: It is assumed that the depleted material is the remnant of \xehs enrichment but  still contains tiny amounts of \xehs. Its abundance is considered to be 0.01\% (scenario I), 0.05\% (scenario II) and 0.1\% (scenario III), respectively. The abundances of each isotope were calculated accordingly and are given in Table~\ref{tab:atoms}. For the studies we used the new precise \zbb half-live value of \tzn = 2.165 $\pm$ 0.016 (stat.) $\pm$ 0.059 (sys.) $\times$ 10$^{21}$ years as measured by EXO-200 \cite{alb14a}.

\begin{table}
\begin{center}
\caption{\label{tab:atoms}
Assumed abundancies (in \%) for the double beta emitters in depleted Xe. \xehf is also given which will be discussed in section 4.}
\footnotesize\rm
\begin{tabular*}{0.7\textwidth}{@{}l*{15}{@{\extracolsep{0pt plus12pt}}l}}

\hline
 & \xehv atoms (\%) & \xehs atoms (\%) & \xehf atoms (\%)\\
\hline
scenario I &  10.5637 & 0.01 & 0.11083\\
scenario II & 10.5629 & 0.05 & 0.11079\\
scenario III & 10.5619 & 0.1  & 0.110727\\

\hline

\end{tabular*}
\end{center}
\end{table}

One of the unknown parameters is the expected half-live of \xehv. Two different types of estimates were performed. First of all, the phase space for \zbb scales with $Q^{11}$. Using a more accurate polynomial given in \cite{boe92} and by assuming equal nuclear matrix elements for the decay mode of \xehv and \xehs with the given experimental \tzn of \xehs \cite{alb14a} the estimate results in an expected value of \tzn $\approx 1.71 \times 10^{25}$ yr. A second approach is using the calculations  \cite{bar13} done with the IBM-2 model and using phase space calculations from the method described in \cite{kot12}. Allowing various values for g$_A$ between 1 and 1.269 with the given matrix elements \cite{iac12} half-lives in the region of 3.7-4.7 $\times 10^{24}$ yrs are predicted. Furthermore, a half-live estimate of 6.09 $\times 10^{24}$\,yr is given in \cite{sta90} which is in the same region.

\subsection{Simulation}
\label{sec:Sim}
\begin{figure}[h]
\centering
\includegraphics[width=0.65\columnwidth]{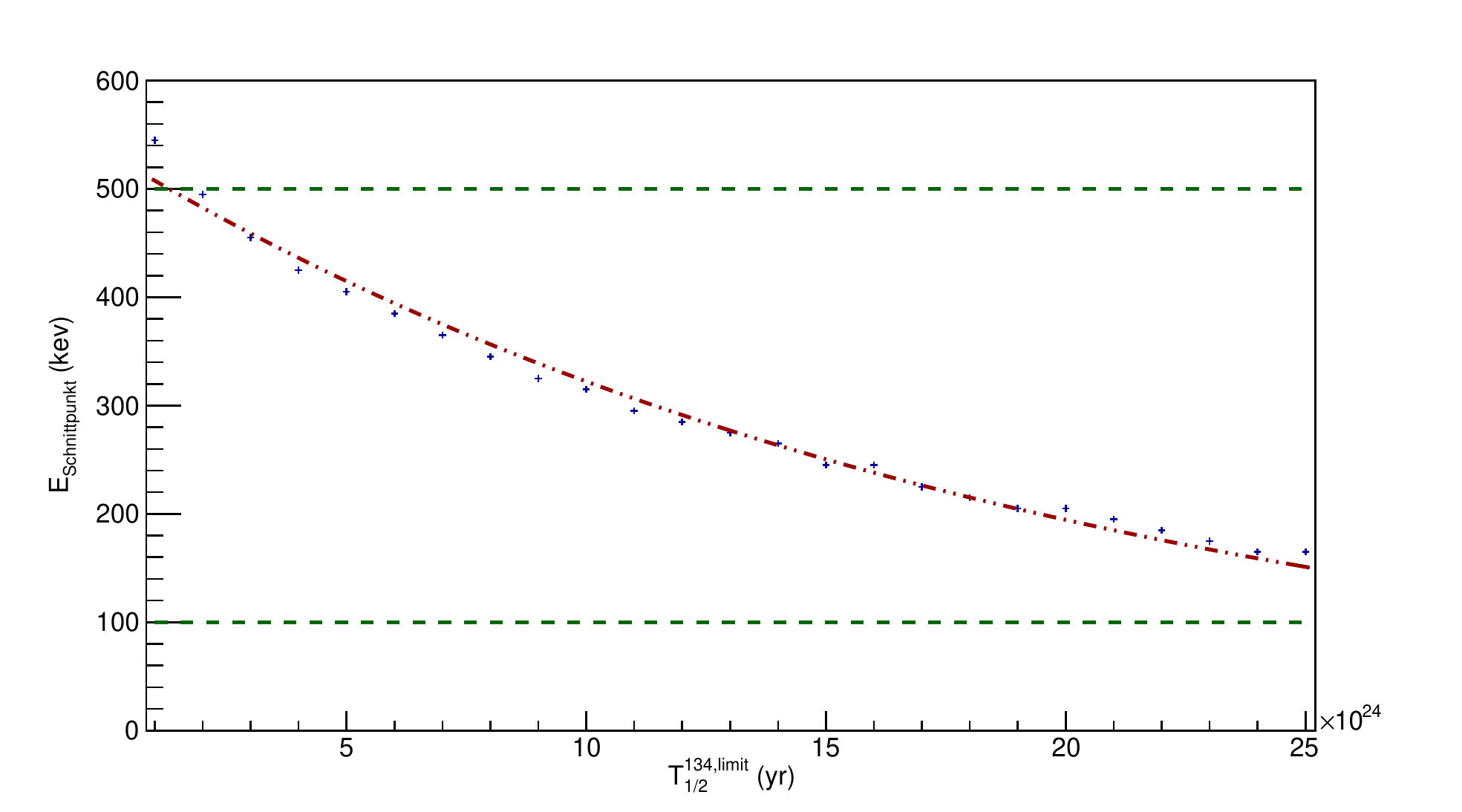}
\caption{The energy for a signal to background ratio of one, i.e. where signal (\xehv) and background (\xehs) have the same value,
as function of the assumed \xehv half-live. For longer half-lives this point moves towards lower energies.}
\label{pic:xpoint}
\end{figure}

To estimate the sensitivity of a large scale LXe experiment, Monte Carlo simulations based on GEANT4 have been performed. The design of the detector is guided by the DARWIN concept using 20 tonnes of LXe and a detector system similar to the one presented  in \cite{obe12} is used. This geometry was implemented into the MaGe simulation software \cite{mag11}. The veto systems for external backgrounds are of no importance for this study and therefore are not discussed. Instead the focus will be on the cryostat with the active volume of LXe. The cryostat is represented by a cylinder with an inner radius of 95.1 cm and a height of 238.7 cm, resulting in a total volume of 6.77 $m^3$. The cryostat contains two distinct volumes filled with xenon: the upper part of the cryostat contains a gas phase with a height of 3.05 cm. The remainder of the cryostat is filled with the liquid phase. At the top and bottom of the cryostat, 500 PMTs are placed to read out the scintillation light. However, within the context of this study, the PMT readout was not simulated, using  instead the deposited energy in the liquid phase. In order to generate the decays of the xenon isotopes of interest, a modified version of the DECAY0 software \cite{pon01}  was used, as the original version of the software does not contain any decay data for other isotopes of interest for this work besides ${}^{136}$Xe. The software was extended to also include the $2\nu\beta\beta$ decay of ${}^{134}$Xe \cite{sim02}, as well as the ECEC, $\beta^{+}$EC and $\beta^{+}\beta^{+}$ of ${}^{124}$Xe \cite{nes12,kim83}. 
As mentioned, the \zbb double beta spectrum typically peaks around 30 \% of the Q-value. For \xehv a significant contribution of the betas is expected to be in the region around 200 keV. The major background for the measurement of this decay will be ${}^{136}$Xe and \kraf, whose half-lives are well known. Therefore, it is of interest to check what is the best energy region to perform the study, as a function of the half-life of \xehv. These results are shown in Fig.~\ref{pic:xpoint}. As expected, the location in energy where a similar amount of events from \xehv and \xehs occurs, varies with the half-life of  \xehv, with longer half-lives having an ideal analysis range towards lower energy values.

\begin{figure}[h]
\centering
\includegraphics[width=0.65\columnwidth]{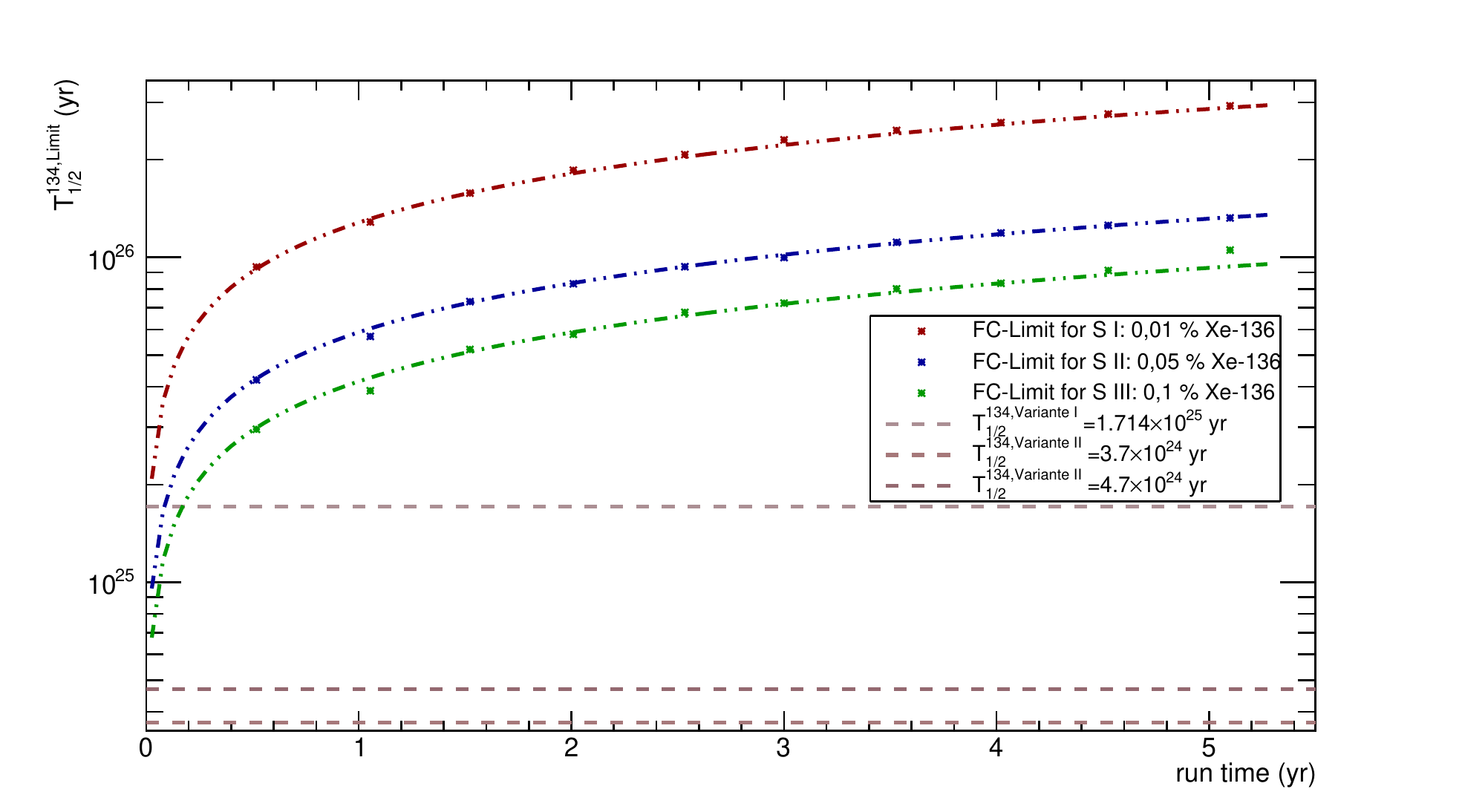}
\caption{Sensitivity for the search of \zbb of \xehv as a function of measuring time in the most optimal case of \xehs being the only source of background. 
In this perfect scenario even
the most conservative approach for the half-life will be reached within 2 years.Horizontal lines represent different \zbb half-lives of \xehv as described in 
section \ref{sec:estimates}.}
\label{pic:sensi}
\end{figure}

As an initial feasibility test the optimal scenario I has been used, assuming a measuring time of 5 years. For a given half-life of 5 $\times 10^{24}$ yr (a very conservative upper bound on the expected half-live using the IBM-2 matrix elements) the expected spectrum can be clearly identified. Similar studies have been performed for all 3 scenarios and various half-lives. All these scenarios demonstrated to be possible to detect the \xehv \zbb, however with different significances.
 To explore the real sensitivity a Monte Carlo study using Poisson distributions for potential background fluctuations
 was performed using the Feldman-Cousins approach \cite{fel98}. No fiducial volume cut was applied. The resulting curves presented in Fig.~\ref{pic:sensi} show that in 5 years measuring time have-lives above 10$^{26}$ years should be accessible. This is well beyond even the most conservative assumption made, however it is assuming no 
 additional background components. 

\newpage
\section{Study of \xehv including solar pp neutrinos}
\begin{figure}[h]
\centering
\includegraphics[width=0.7\columnwidth]{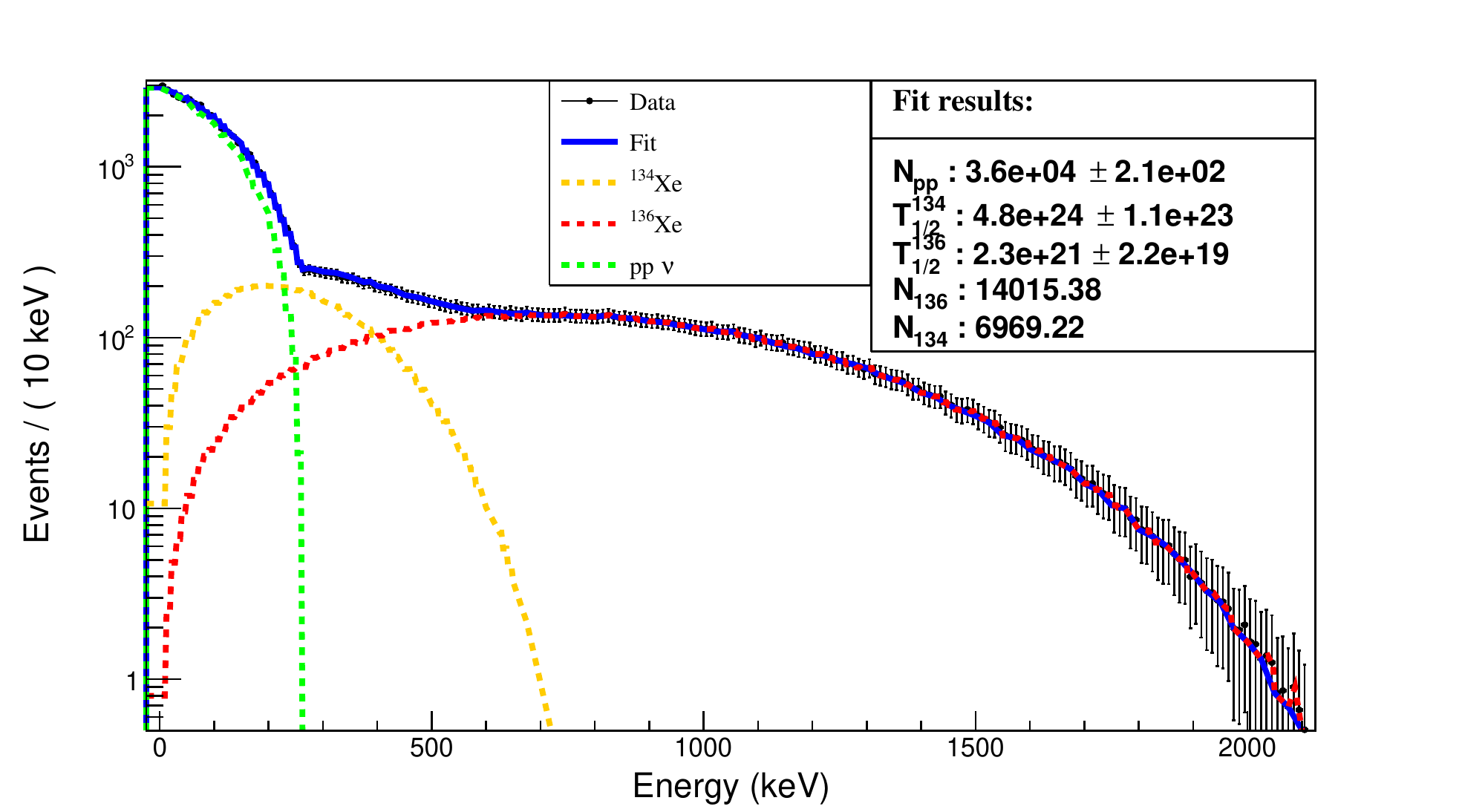}
\caption{A fit based on an extended maximum likelihood approach using a half-life of 4.8$\times 10^{24}$ yrs for \xehv, the \xehs \zbb half-life and the recoil spectrum from neutrino-electron scattering by solar pp-neutrinos. In the range of about 250-450 keV the \xehv signal is still clearly visible.}
\label{pic:mlfit}
\end{figure}

The discussed optimal scenario will be worsened by potential background components like radioactive impurities such as $^{85}$Kr, or 
electron recoils from scattering of solar neutrinos. Here the focus is on the electron recoils from solar pp neutrinos. The endpoint 
of this recoil spectrum is at 260 keV, hence effecting the window for the \xehv search from the low energy side. 
For the scenario III the corresponding simulated spectrum is shown for 4.8 $\times 10^{24}$ yr in Fig.~\ref{pic:mlfit}. 

\begin{figure}[h]
\centering
\includegraphics[width=0.65\columnwidth]{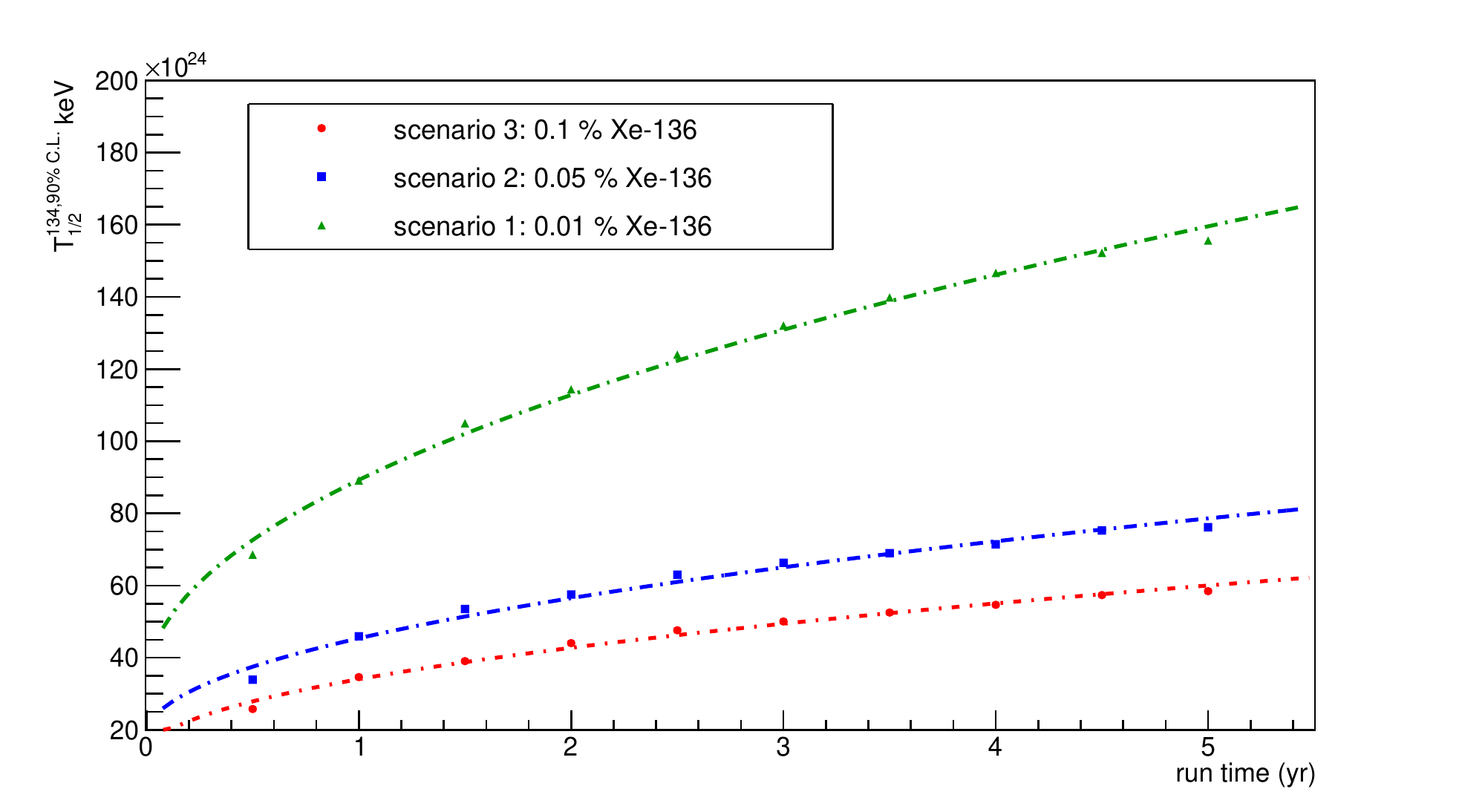}
\caption{Sensitivity for the search of \zbb \ of \xehv as a function of measuring time including solar neutrino background.}
\label{pic:sense}
\end{figure}

A sensitivity study has been performed with the extended maximum likelihood method described in \cite{bar90}. The deduced sensitivity curves are shown in  Fig.~\ref{pic:sense}. 
As can be seen, still the most conservative estimate from Table \ref{tab:theory} can be reached, however it is not possible anymore to measure a half-life
beyond about 2$\times$10$^{26}$ yrs in 5 years.

\section{Study of \xehv including \kraf}
\begin{figure}[h]
\centering
\includegraphics[width=0.75\columnwidth]{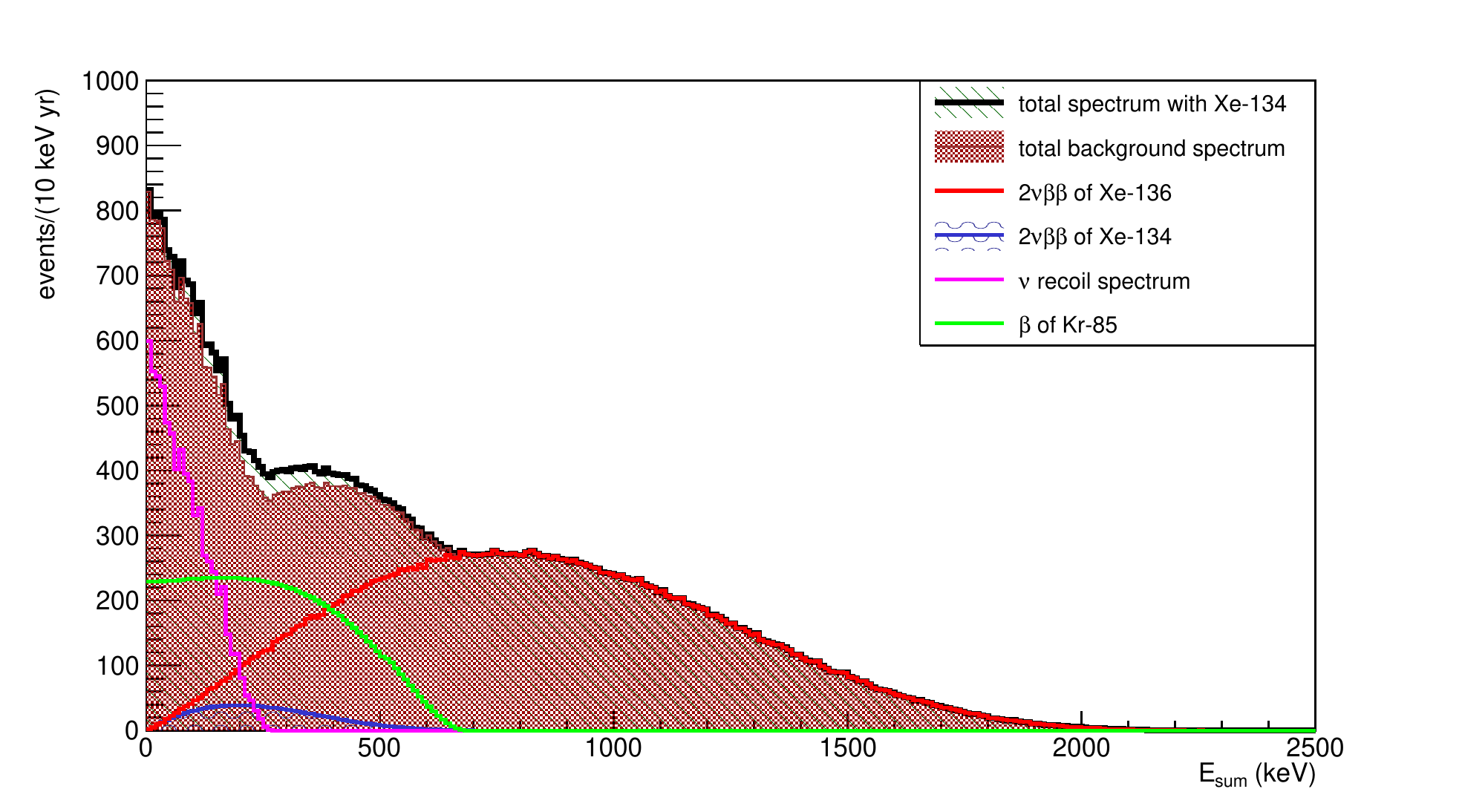}
\caption{Background model including solar neutrino background,\xehs and \kraf with an assumed purification level of 0.1\,ppt Kr/Xe. The assumed
\zbb half-live of \xehv is 5 $\times$ 10$^{24}$\,yr.}
\label{pic:backgroundmodelWithKr}
\end{figure}

\begin{figure}[h]
\centering
\includegraphics[width=0.75\columnwidth]{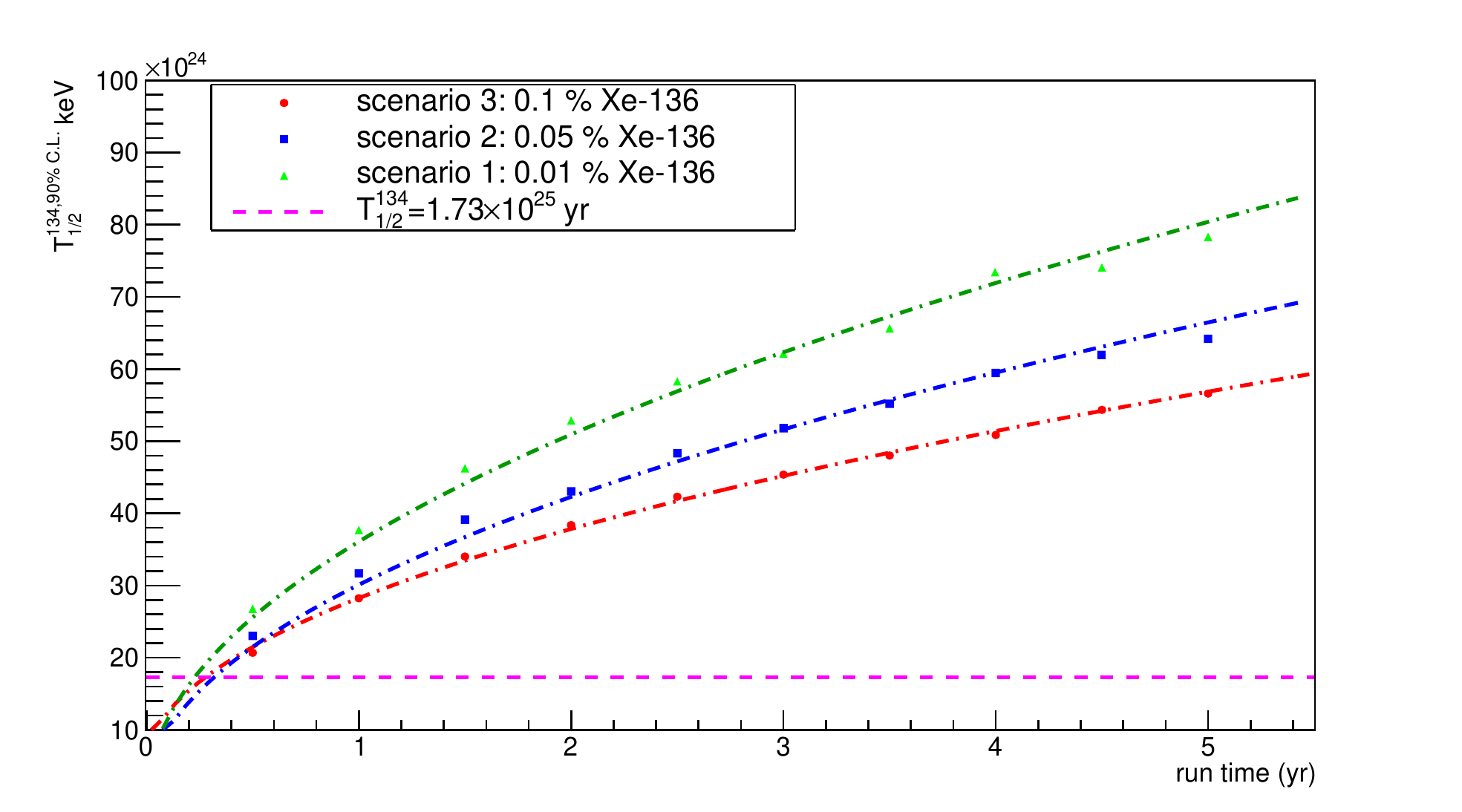}
\caption{Sensitivity for the search of \zbb of \xehv as a function of measuring time including solar neutrino background and the decay of \kraf \ with an assumed purification level  of 0.1\,ppt Kr/Xe.}
\label{pic:sensiWithKr}
\end{figure}

Xenon gas obtained from the atmosphere to be used in LXe-TPCs contains a variety of traces from other elements, which provides additional background in the experiment. 
Most of these backgrounds can be reduced, by different purification methods, to a level where they are no longer relevant for the experiment.
One of such impurities is krypton which, being also a noble gas, it particularly hard to separate from the xenon.  One of the krypton isotopes, \kraf, has an abundance of $2\times10^{-11}$\cite{xdu04}, which decays by beta emission with a Q-value of 687.4\,keV and a half-life of 10.76\,yrs.
The spectral shape of the beta spectrum was taken from DECAY0.  Adding this single-unique forbidden beta decay to the background model increases the expected number of events in the region of interest for \xehv, as can be seen in (Fig.~\ref{pic:backgroundmodelWithKr}).
As krypton is evenly distributed in xenon the amount of events resulting from this decay increases with the size of the detector and can not be reduced through fiducial volume cuts.
One way of reducing this background is through purification to reduce the levels of Kr/Xe. In this work, it is assumed that a purification procedure is carried out on reaching a \kraf abundance of 0.1 ppt (Fig.~\ref{pic:backgroundmodelWithKr}), which is a factor 5 below the 0.5 ppt goal of the XENON1T-collaboration.
However, it is a reasonable guess that with the experience from previous experiments the \kraf-level for a multi-tonne detector 
can be improved by this factor. 
However, this background might decrease over the measurement period. Even using a reasonable assumption, this choice of 0.1\,ppt is still arbitrary, which prompts for further studies to be carried on.
The resulting sensitivity for \xehv is shown in (Fig.~\ref{pic:sensiWithKr}). The introduction of \kraf into the analysis shows a reduction of the sensitivity by a factor of two with the chosen purification level.

\section{Study of \xehf and \xehz}

Another isotope of interest would be \xehf, which is one out of only six nuclides able to perform
double positron decay. This implies additional competing decay modes containing also electron capture (EC)  and hence very different signatures for all decay modes (see Eq. \ref{eqn:b+b+}$-$\ref{eqn:ecb+}). They will be discussed separately. Topological information has not been included in these studies and cannot be done in LXe, however the NEXT experiment has published  results about the topological information from a gaseous Xe-detector \cite{alv13}.

\begin{table}[h]
\caption{\label{tab:theory}
Calculated half-lives for the three different 2$\nu$ decay modes of \xehf.}
\footnotesize\rm
\begin{tabular*}{\textwidth}{@{}l*{15}{@{\extracolsep{0pt plus12pt}}l}}
\hline
 & $2\nu$ ECEC (10$^{21}$) yr &  $2\nu$ \bec  (10$^{22}$) yr &  $2\nu$ \bpbp (10$^{27}$) yr & Reference\\
\hline
& 2.9-7.3 & 3.0-7.6 & 1.4-3.54 & \cite{hir94}\\
& 390-986 & 230 - 582 & - & \cite{aun96}\\
& 7.0 & 8.2  & -  & \cite{rum98}\\
& 61-155 & 72-182 & 2.6-6.6 & \cite{shu07}\\
& 7.1-18 & 8.3- 21.1 & 3.0 -7.6  & \cite{sin07}\\
& 0.4-8.8 & 0.94-9.7 & 0.17 -3.8 & \cite{suh13}\\
\hline
\end{tabular*}
\end{table}

\subsection{The 2 $\nu$ ECEC case} 
As the full phase space is available for 2$\nu$ ECEC, it is the most likely mode to occur. Expected half-lives
are in the same order as for the \zbb and theoretical predictions are compiled in Tab.~\ref{tab:theory}. As can be seen a wide spread of calculated half-lives exists. 

\begin{figure}[h]
\centering
\includegraphics[width=0.75\columnwidth]{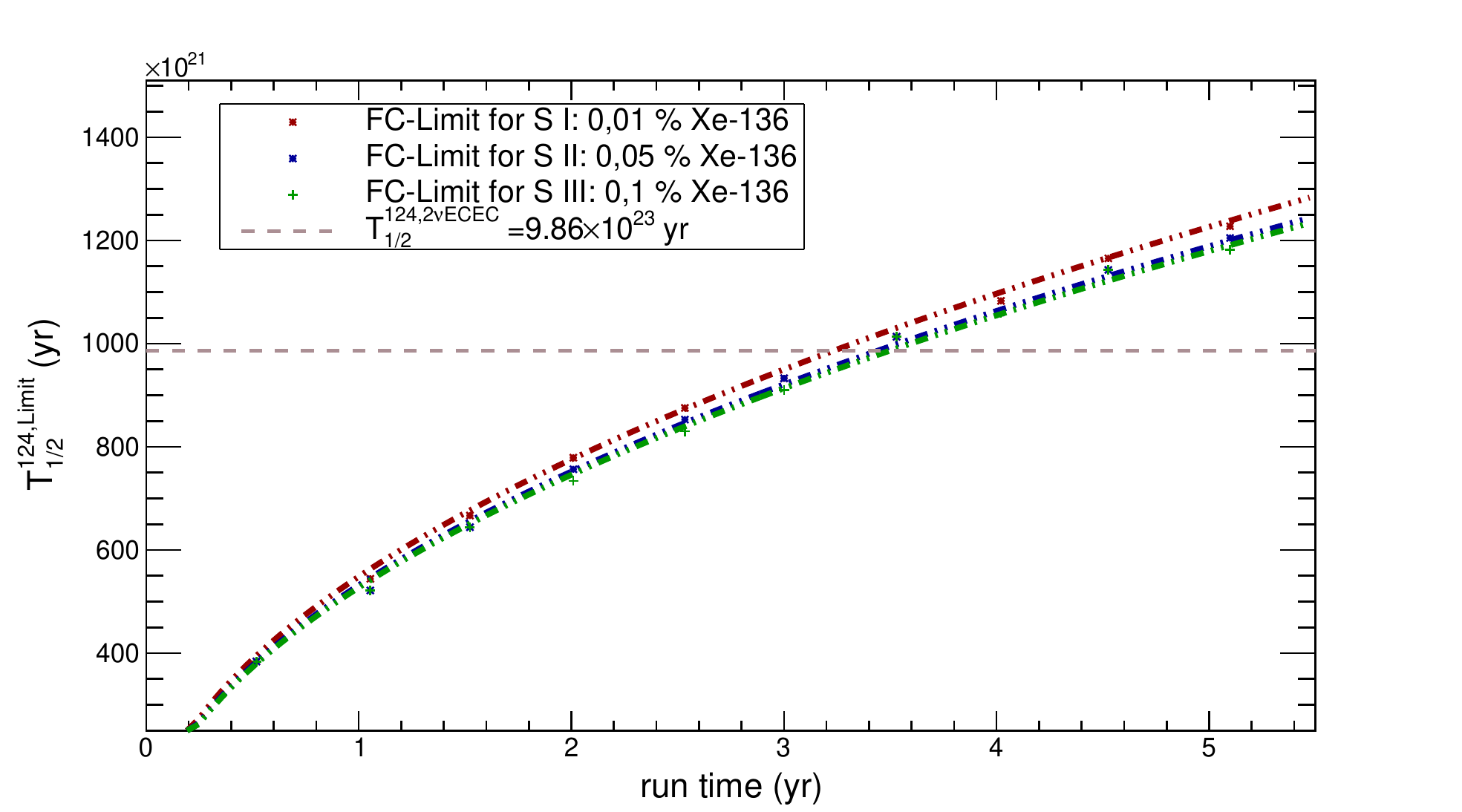}
\caption{Feldmann - Cousins sensitivity on the \xehf double EC half-life as a function of running time.
The region considered is 10-100 keV.}
\label{pic:ECEC}
\end{figure}

The predictions for 2$\nu$ ECEC half-lives span a range of two orders of magnitude, experimentally a lower half-life limit for
the \xehf decay of 1.1 $\times 10^{17}$\,yr is given \cite{gav98}. No experimental limit exists for \xehz and only one prediction is available for a half-life value between 4.7-12 $\times 10^{25}$ yrs \cite{shu07}. \xehz has a Q-value of 897 keV and an abundance about the same as \xehf (0.09 \% compared to 0.1\%). The signature in both cases are the same, namely X-rays and Auger electrons of Te from the captures of the inner shells. However, using a crude assumption of equal matrix elements
for both Xe-isotopes due to the $Q^5$ power phase space dependence of 2 $\nu$ ECEC the expected half-life for \xehz will be about 340 times longer than the one for \xehf. Thus, a possible signal will always be dominated by \xehf. \\
The details of the individual X-ray emission lines and their intensities could be lost due to the energy resolution of the detector. Taking an energy resolution of $\sigma (E)/E$ of about 9 \% at 50 keV as measured by Xenon100 \cite{apr12}, a single peak will result at about 55 keV. \\
Major background components will again be solar pp neutrinos, \kraf decay and the \zbb contributions of \xehs and \xehv. The expected sensitivity for a 2$\nu$ ECEC search using the Feldman-Cousins as before approach, except that this time \kraf is not included,
is shown in Fig.~\ref{pic:ECEC}. 

\subsection{ The 2 $\nu$ \bec mode}
\begin{figure}
\centering
\includegraphics[width=0.75\columnwidth]{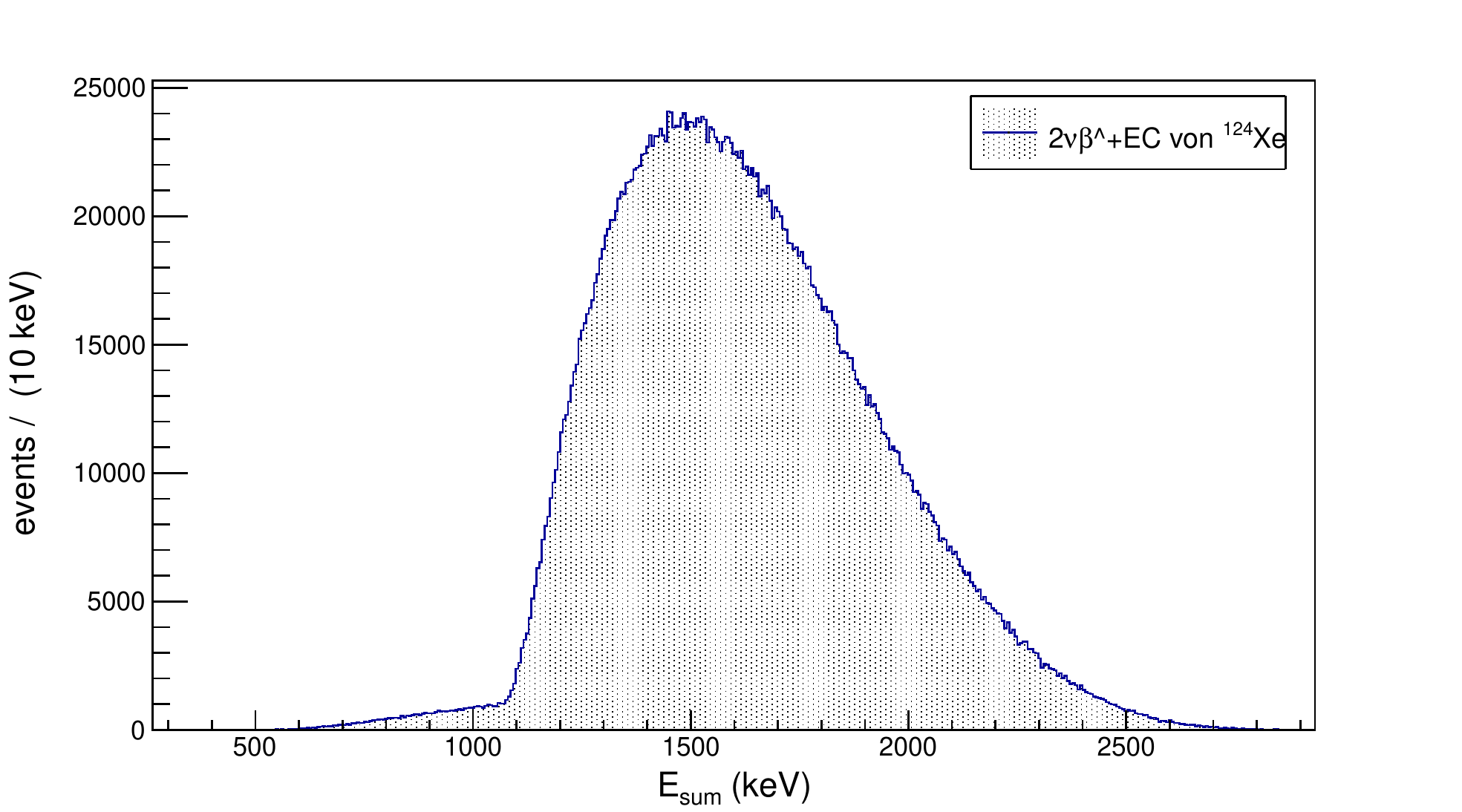}
\caption{Schematic plot of the pure expected 2$\nu$ \bec signal of \xehf decay as produced with a modified DECAY0,
no smearing due to energy resolution is applied.}
\label{pic:bec}
\end{figure}

The signal of this decay mode, only possible for \xehf, is the kinetic energy of the positron
combined with the energy of the annihilation photons and the X-rays and Auger electrons (see Fig.~\ref{pic:bec}) . 
In this way the signal is much higher in energy, producing a continuous spectrum starting
around 2 m$_ec^2$ with the major part occurring between 1-2 MeV. This might require a dedicated data
acquisition system, as the dynamic range of LXe detectors searching for dark matter are not designed
for such high energies.

 Hence, from the background
contributions mentioned above only \xehs matters. Estimations for the half-lives compiled in Tab.~\ref{tab:theory} show
variations over two orders of magnitude, basically from 3$\times 10^{22}$\,-\,3$\times 10^{24}$ yrs while the experimental lower limit is 4.8 $\times 10^{16}$ yr (68 \% CL) \cite{bar89}.

Detailed studies for the three scenarios using the energy resolution of EXO-200 \cite{aug12} show that the full range of theoretical predictions
can only be explored in the scenarios I and II. As scenario III might be the most realistic one, 
the investigation reveals that the longest half-lives predicted might not be accessible.
Note, only energy but no spatial information was used, which might help to identify the signal
using the specific topology resulting from positron annihilation.

\subsection{ The 2 $\nu$ \bpbp mode}
Despite its  signature of four 511 keV photons
combined with the kinetic energy of the two positrons, this mode suffers from a reduced phase space.
Theoretical half-live estimates are all in the region between $10^{27} - 10^{28}$ years \cite{hir94,aun98,sin07}
and thus make a successful search even in a large scale detector as discussed in this paper unlikely 
as event rates are less than one per 10 years, even though this mode will be more or less background free.

\section{Summary and conclusions}
Ongoing and planned large scale LXe experiments for dark matter searches allow investigations for 
various double beta emitters besides \xehs especially in depleted Xe. In this paper the three additional
nuclides \xehf, \xehz and \xehv are discussed. It has been shown that under realistic assumptions
on abundances of nuclides in depleted Xe, presented here as different scenarios, and energy resolution
for a wide range of nuclear matrix element calculations the discovery of neutrino accompanied modes
should be feasible. 

For the \xehv search neutrino - electron scattering from solar neutrinos, \kraf beta decay and the
\zbb of \xehs are dominant backgrounds. The same appears for the $2\nu$ ECEC search with the addition that \zbb of \xehv will
contribute as well. 

The decay modes containing positrons suffer from reduced phase spaces. In the \bec mode only
in the most optimistic scenarios the full range of predicted half-lives can be explored, being more likely that only a fraction of it will be accessible. 
The \bpbp mode might be not observable even in 
such large scale experiments as the predicted half-live is too long. No fiducial 
volume cut was applied to the detector, assuming it is fully active. As background components
only solar pp neutrinos, \kraf and potential double beta decays were considered. Further ones like $^7$Be solar neutrinos and others might 
contribute as well and will be discussed together with the mentioned ones in a separate, more detailed and systematic paper.

\section{Acknowledgement}
The authors would like to thank F. Iachello for providing matrix elements and phase spaces and the GERDA and Majorana collaborations for the permission to use the MaGe simulation software.
This work is supported by the Helmholtz Alliance for Astroparticle Physics.

\bigskip
\noindent After submission we became aware of the preprint \cite{mei13} which is discussing one of the decay modes mentioned
in this paper for a specific smaller scale experiment.


\section*{References}
\begin{thebibliography}{10}
\bibitem{fon12} C.~S.~Fong, E.~Nardi, A.~Riotto, Adv. High Energy Phys. 2012, 158302 (2012)
\bibitem{gan13} A.~Gando et al.,  \Journal{\PRL}{110}{062502}{2013}
\bibitem{ago13} M.~Agostini et al., \Journal{\PRL}{111}{122503}{2013}
\bibitem{alb14} J.~B.~Albert et al., \Journal{\NAT}{510}{229}{2014}
\bibitem{bes10} D.~R~ Bes, O.~ Civitarese,  \Journal{\PRC}{81}{014315}{2010}
\bibitem{hir94} M.~ Hirsch et al., \Journal{\ZPA}{347}{151}{1994}
\bibitem{apr12} E.~Aprile et al.,  \Journal{\APPH}{35}{573}{2012}
\bibitem{ake14} D.~S.~Akerib et al.,  \Journal{\PRL}{112}{091303}{2014}
\bibitem{abe13} K.~Abe et al., \Journal{\NIMA}{716}{78}{2013}
\bibitem{mal11} D.~C.~Malling et al., arXiv:1110.0103
\bibitem{bau11} L.~Baudis et al., \Journal{\JPCS}{375}{012028}{2012}
\bibitem{ame12}  M.~Wang et al., \Journal{\CPC}{36}{1603}{2012}
\bibitem{bar89} A.~Barabash et al.,  \Journal{\PLB}{223}{273}{1989}
\bibitem{red07} M.~Redshaw et al.,  \Journal{\PRL}{98}{053003}{2007}
\bibitem{aug12} M.~Auger et al.,  \Journal{\PRL}{109}{032505}{2012}
\bibitem{gan12} A.~Gando et al.,  \Journal{\PRC}{85}{045504}{2012}
\bibitem{alb14a} J.~B.~Albert et al., \Journal{\PRC}{89}{015502}{2014}
\bibitem{boe92} F.~Boehm, P.~Vogel, Physics of massive neutrinos, Cambridge Univ. Press 1992
\bibitem{bar13} J.~ Barea, J.~ Kotila and F. ~Iachello,  \Journal{\PRC}{87}{014315}{2013}
\bibitem{kot12} J.~ Kotila and F.~Iachello,  \Journal{\PRC}{85}{034316}{2012}
\bibitem{iac12} F.~Iachello,  private communication
\bibitem{sta90} A.~Staudt, K.~Muto, H.~V.~Klapdor-Kleingrothaus, \Journal{\EPL}{13}{31}{1990}
\bibitem{obe12} U.~Oberlack, talk presented at 8th PATRAS workshop on axions, WIMPs and
WISPs, 2012
\bibitem{mag11}M.~Boswell, et al., IEEE Trans.Nucl.Sci., vol. 58, no. 3, pp. 1212-1220 (2011)
\bibitem{pon01} O.~A.~Ponkratenko, V.~I.~Tretyak, Y.~U.~Zdesenko, \Journal{\PAN}{63}{1282}{2000}
\bibitem{sim02}F.~Simkovic, P.~ Domin, A. ~Faessler, arXiV:hep-ph/0204278v1(2002)
\bibitem{nes12} D. A. Nesterenko et al., \Journal{\PRC}{86}{044313}{2012}
\bibitem{kim83}C.~Kim,  K.~Kubodera, \Journal{\PRD}{27}{2765}{1983}
\bibitem{fel98} G.~J.~Feldman,  R.~D.~Cousins,  \Journal{\PRD}{57}{3873}{1998}
\bibitem{bar90} R.~Barlow,  \Journal{\NIMA}{297}{496}{1990}
\bibitem{xdu04} X. ~Du et al. \Journal{\RoSI}{75}{10}{2004}
\bibitem{aun96} M.~ Aunola, J.~Suhonen \Journal{\NPA}{602}{133}{1996}
\bibitem{rum98} O. ~A.~ Rumyantsev, M.H. Urin \Journal{\PLB}{443}{51}{1998}
\bibitem{alv13} V.~Alvarez et al.,  \Journal{\JINST}{8}{P09011}{2013}
\bibitem{gav98} J.~ M.~ Gavriljuk et al. \Journal{\PAN}{61}{1287}{1998}
\bibitem{sin07} S.~Singh et al.,   \Journal{\EPJA}{33}{375}{2007} 
\bibitem{aun98} M. Aunola, J. Suhonen \Journal{\NPA}{643}{207}{1998}
\bibitem{shu07} A.~ Shukla, P.~K.~ Raina , P. ~K.~ Rath \Journal{\JPG}{34}{549}{2007}
\bibitem{suh13} J. ~Suhonen  \Journal{\JPG}{40}{075102}{2013}
\bibitem{mei13} D.~M.~ Mei et al., arXiV:1310.1946


\end{thebibliography}
\end{document}